\documentclass[fleqn,12pt,twoside]{article}
\usepackage{espcrc1}

\usepackage{graphicx}
\usepackage[figuresright]{rotating}

\newcommand{\AmS}{{\protect\the\textfont2
  A\kern-.1667em\lower.5ex\hbox{M}\kern-.125emS}}

\title{New Class of Compact Stars at High Density\thanks{Presented at 
        the International Conference on Statistical QCD, Bielefeld, 
        Germany, 26-30 Aug 2001.}}

\author{E. S. Fraga\address[BNL]{Department of Physics, Brookhaven 
        National Laboratory, \\ Upton, NY 11973-5000, USA}%
        \thanks{Present address: LPT, Universit\'e Paris-Sud XI, B\^atiment
        210, F-91405 Orsay, France.}
        R. D. Pisarski\addressmark[BNL]
        J. Schaffner-Bielich\address{RIKEN BNL Research Center, 
        Brookhaven National Laboratory, 
        Upton, NY 11973-5000, USA}}
       
\begin{document}

\maketitle

\begin{abstract}
We discuss the equation of state for cold, dense quark
matter in perturbation theory, and how it might match
onto that of hadronic matter.  Certain choices 
of the renormalization scale correspond to a strongly first order chiral 
transition, and may generate a new class of small and very dense quark stars. 
The results for the mass-radius relation are compatible with the recent 
determination of the mass and the radius of an isolated neutron star
by Pons {\it et al}.
\end{abstract}
\vspace{1cm}

The investigation of the structure of the phase diagram for 
strongly interacting matter is a fascinating subject.
Moreover, the study of regions of the diagram associated with 
extreme conditions in temperature and density can reveal new 
phenomena, and provide a better understanding of Quantum Chromodynamics (QCD). 
Nevertheless, in spite of the increasing interest this field has 
attracted during the last decade, the mapping of the QCD phase 
diagram is still in its infancy, and most of its regions remain 
mystifying and challenging. An exception is given by the region of 
nonzero temperature and very small densities. There, guidance is 
given by lattice simulations \cite{lattice} 
and experiments in heavy ion collisions \cite{RHIC}. 
On the other hand, an increasing amount of precise astrophysical 
data provides a new resource to probe QCD at 
large density. The natural candidates to study are 
neutron stars, whose interior might be dense enough to allow
for the presence of chirally symmetric quark matter 
\cite{G_b,Free,Baluni,strange,thirda,thirdb}. 
Concretely, certain combinations of observables, such as the mass-radius 
relation, may be useful to constrain the equation of state for strongly 
interacting matter.

Neutron star models predict a maximum mass 
in the range $\approx 1.4 $ -- $ 2.0 M_\odot$, with a radius 
$\approx 10$ -- $15$~km. The values for the maximum mass 
are consistent with the pulsar data \cite{G_b}. The conventional 
theoretical approach used to describe 
a hybrid star, a neutron star with quark matter in the core, or a strange star
is the bag model, with at best a correction $\sim \alpha_s$ from
perturbative QCD \cite{strange}. If the bag constant is fit from
hadronic phenomenology, then the gross features of those stars are very
similar to those expected for neutron stars, especially in the region 
of mass compatible with the pulsar observations. Through the years, the 
study of compact hybrid stars has produced a myriad of different models, with 
all sorts of possible layers and condensates, but with numbers very 
similar to the ones presented above \cite{G_b}. 
Nevertheless, recent astrophysical 
results for the mass-radius diagram seem to point in a direction 
unforeseen by traditional models \cite{Pons:2001px}.

In \cite{Fraga:2001id} we considered cold, dense quark matter in 
perturbation theory to $\sim \alpha_s^2$ \cite{Free,Baluni},
analyzing its implications for the physics of quark stars.
We used perturbation theory not because it is a 
good approximation, but simply to model the possible nonideality in 
the equation of state of cold, dense QCD.
Here we extend this discussion, and consider 
as well how to match this ``perturbative'' equation of state 
onto that of hadrons.  We emphasize especially the comparison between
the pressure of the hadronic and quark phases.  While our results
are admittedly similar to other theoretical approaches
\cite{Peshier:2001pn}, we try to place these results in a more
general context, in terms of the QCD phase diagram.
Lastly, most intriguing is the 
recent determination of the mass and the radius 
of an isolated neutron star, which suggests an ``implausible''
equation of state \cite{Pons:2001px}.

Assume that the chiral phase transition occurs at a chemical
potential $\mu_\chi$. Our perturbative equation of state applies
in the chirally symmetric phase, when the quark
chemical potential $\mu > \mu_\chi$.
In this phase, the effects of a strange quark mass,
$m_s\approx 100$~MeV \cite{Blum99}, are small relative to the
quark chemical potentials, $\mu > 300$~MeV.  Thus
we take three flavors of massless quarks with equal
chemical potentials \cite{G_b,raj}.  
How the equation of state for quarks matches onto that of hadrons
is discussed later.

The thermodynamic potential of a plasma of massless quarks 
and gluons was calculated perturbatively to $\sim \alpha_s^2$
by Freedman and McLerran \cite{Free} and by Baluni \cite{Baluni},
using the momentum-space subtraction (MOM) scheme.
The MOM coupling is related to that in the
modified minimal subtraction scheme, $\overline{{\rm MS}}$, as
\cite{Fraga:2001id,explanation}
\begin{equation}
\frac{\alpha_s^{{\rm MOM}}}{\pi}=\frac{\alpha_s^{\overline{{\rm MS}}}}{\pi}\;
\left[ 1+{\cal A}~\frac{\alpha_s^{\overline{{\rm MS}}}}{\pi}\right] \quad ;
\label{e1}
\end{equation}
$\alpha_s = g^2/(4 \pi)$, with $g$ the QCD coupling constant,
and ${\cal A}=151/48 - (5/18) N_f$, with $N_f$ the number
of massless flavors. In the $\overline{{\rm MS}}$ scheme, 
to $\sim \alpha_s^2$ the thermodynamic potential is \cite{Fraga:2001id}
\begin{equation}
\label{e2}
\Omega(\mu)= - \frac{N_f \mu^4}{4\pi^2}
\left\{1-2 \left(\frac{\alpha_s}{\pi}\right) -
\left[G+N_f\ln{\frac{\alpha_s}{\pi}}
+ \left(11-\frac{2}{3} N_f \right) \ln{\frac{\bar\Lambda}{\mu}} \right]
\left(\frac{\alpha_s}{\pi}\right)^2 \right\} \; ,
\label{eq:omega}
\end{equation}
where $G=G_0-0.536N_f+ N_f\ln{N_f}$, $G_0=10.374 \pm .13$, and
$\bar\Lambda$ is the renormalization subtraction point. 
We take the scale dependence of the strong coupling constant,
$\alpha_s \equiv \alpha_s(\bar\Lambda)$ to three loop order \cite{PDG}.
In $\overline{{\rm MS}}$ scheme, the thermodynamic potential is
manifestly gauge invariant. However, there remains a dependence on 
$\bar\Lambda/\mu$ that, in principle, can be freely chosen. Although the 
results are significantly sensitive to the particular value of 
this free parameter, this choice is tightly constrained by the physics. 
(It is not especially sensitive to what order in the loop expansion
one used for the running of the coupling constant, or for the pressure.)

With this ``perturbative'' equation of state --- used far outside
the region of validity for perturbation theory ---
reasonable values turn out to be in the range $2\leq  \bar\Lambda/\mu\leq 3$. 
This comes about from the most elementary considerations of how
to match the quark and hadronic equations of state.
\begin{figure}[htb]
\begin{minipage}[t]{71mm}
\includegraphics[width=7cm]{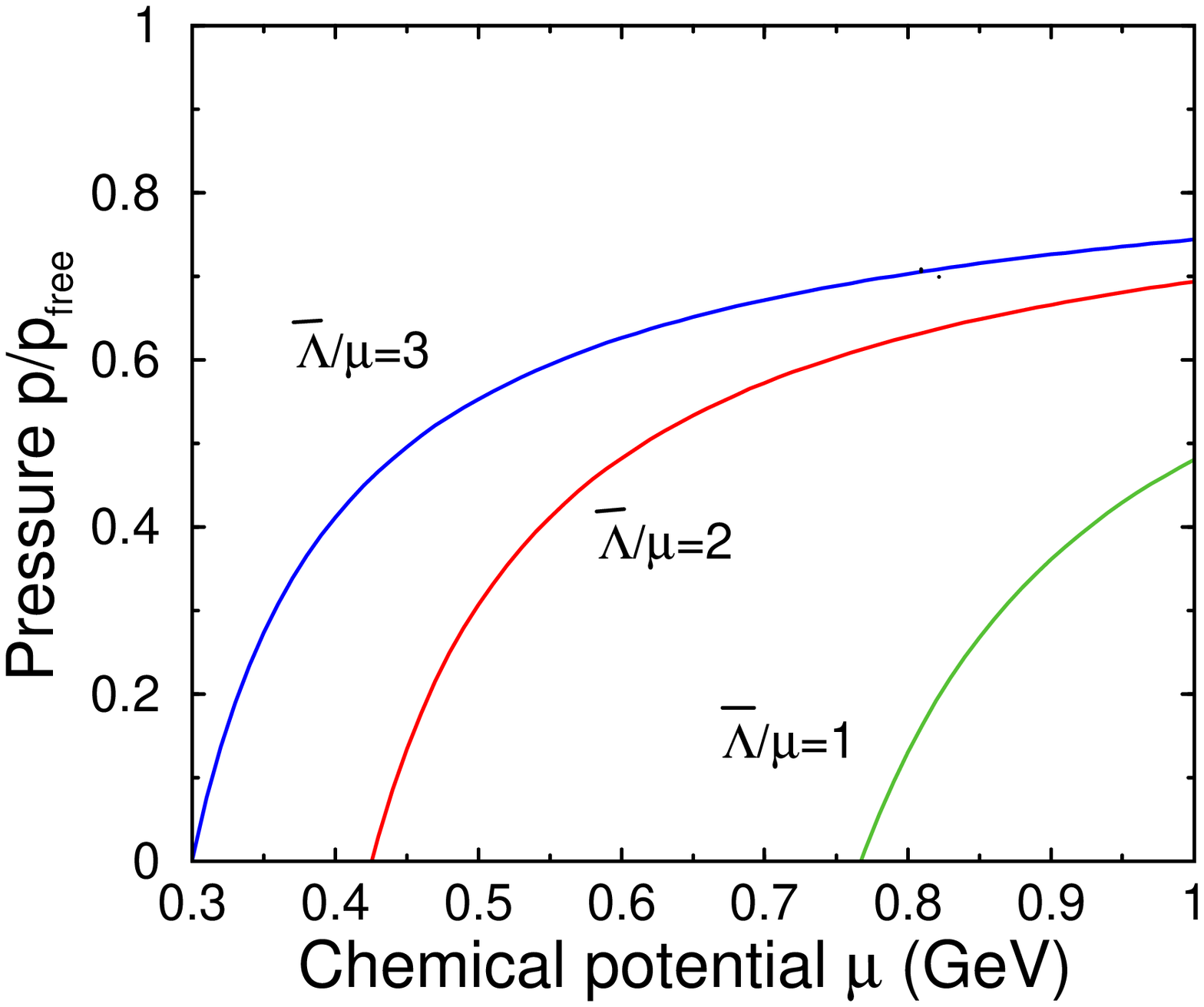}
\caption{The total pressure up to order $\sim \alpha_s^2$, 
relative to the pressure of an ideal gas, $p_{free}$, as a function 
of $\mu$ for different choices of $\bar\Lambda/\mu$.}
\label{fig:pmuq_123}
\end{minipage}
\hspace{\fill}
\begin{minipage}[t]{71mm}
\includegraphics[width=7cm]{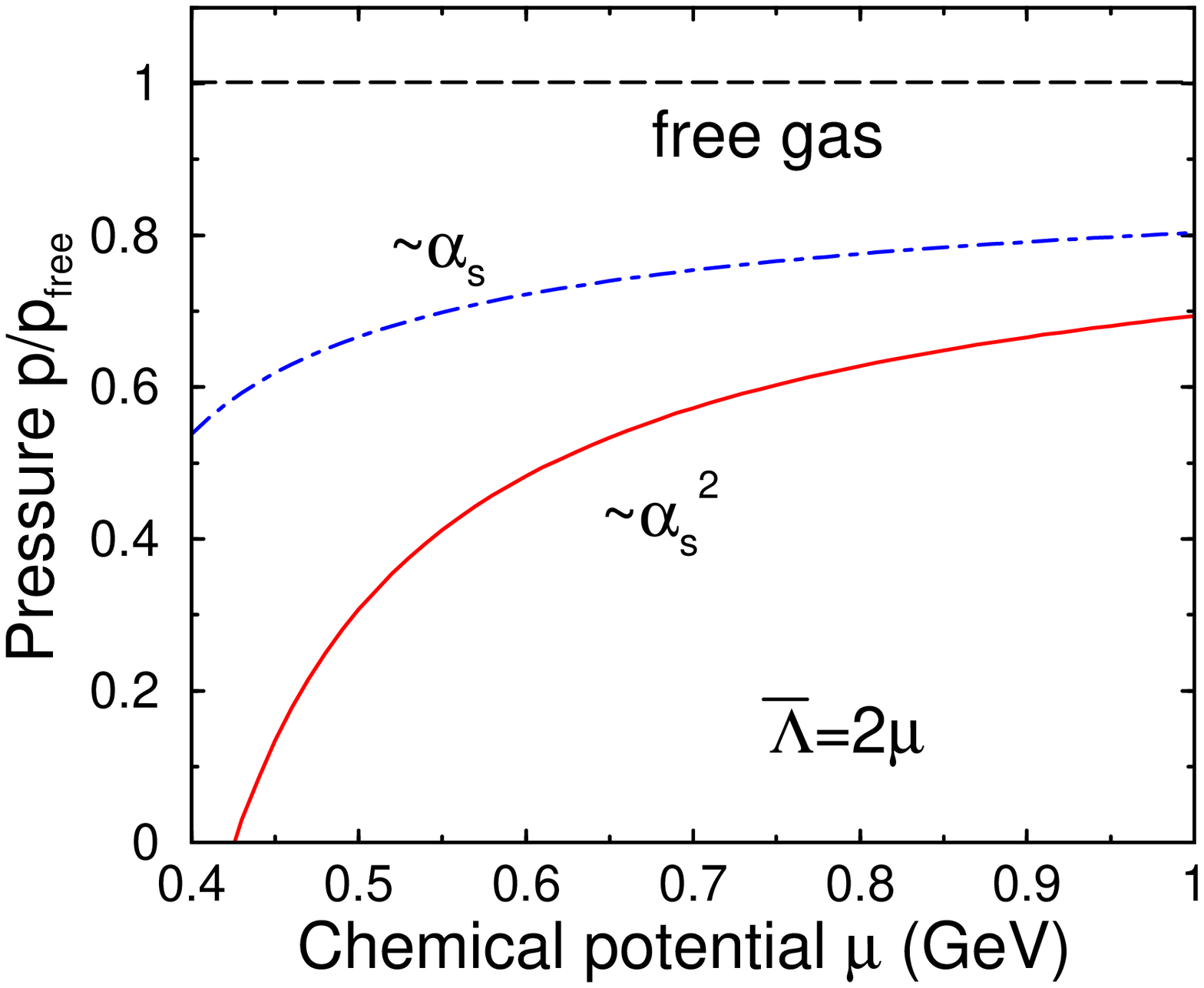}
\caption{The total pressure, relative to the pressure of an
ideal gas, $p_{free}$; including terms to order
$\sim \alpha_s$ and to order $\sim \alpha_s^2$, as a function 
of $\mu$; $\bar\Lambda=2\mu$.}
\label{fig:pmuq2}
\end{minipage}
\end{figure}

In Fig. \ref{fig:pmuq_123} we show
the pressure for some illustrative choices of 
$\bar\Lambda/\mu=1,2,3$.  
In each case, the pressure vanishes at some
$\mu=\mu_c$.  As the chemical potential
of any field must be greater than the mass, and as
the chemical potential of quarks is one third that of baryons,
the hadronic pressure vanishes at $\mu_{min} = 939/3=313$~MeV 
(This neglects the small effects of nuclear binding,
which contribute at most $\sim 15/3=5$~MeV).
Then $\mu_c$ must be greater than $\mu_{min}$; 
from Fig. \ref{fig:pmuq_123}, 
this requires that $\bar\Lambda$ is less 
than $3\mu$.  Conversely, $\mu_c$ cannot be too large.
Dense hadrons exert pressure, so before the quark pressure
goes negative, a transition to massive quarks and hadronic matter occurs.
For example, when $\bar\Lambda=\mu$, one can show that hadronic matter
must persist to very high densities, maybe thirty times nuclear
matter density.  This is unlikely; indeed, below we see that the worry is not
how small the hadronic pressure is, but how big.

Thus, reasonable values of $\bar\Lambda$ are in the range of
$\mu\rightarrow3\mu$.  Choosing the value $\bar\Lambda = 2\mu$,
we can get an idea of how well perturbation theory works
for large chemical potential.
At $\mu=1$~GeV, where $\alpha_s=0.31$, the perturbative
corrections to ideality are $\sim 30\%$ ;
at $\mu=100$~GeV, where $\alpha_s=0.095$, they are still $7\%$.
Even so, as we argued in Ref. \cite{Fraga:2001id}, the perturbative series 
at $T=0$ and $\mu\neq 0$ {\em may} be much better behaved than at $T\neq 0$
and $\mu=0$ \cite{Rob00}.
While for $T\neq 0$, $\mu=0$ we have an expansion in $\sqrt{\alpha_s}$, 
for $T=0$, $\mu\neq 0$ it is a perturbative series in 
$\alpha_s$ and $\alpha_s \log(\alpha_s)$. 
Besides, at nonzero density (and $T=0$), the $\sim \alpha_s$,
and $\sim \alpha_s^2$ corrections both act to lower the pressure.
It is encouraging that 
the $\sim \alpha_s^2$ corrections are smaller than those at
$\sim \alpha_s$ down to $\mu = 0.54$~GeV. 
In contrast,
at nonzero temperature (and $\mu = 0$), terms oscillate in sign,
and the naive perturbative series is not well behaved until 
extremely high temperatures, $\sim 10^7$~GeV\cite{braaten}.

Of course a much better estimate of the quark pressure
could be obtained if the terms 
$\sim \alpha_s^3(\log^2(\alpha_s),\log(\alpha_s),1)$
were computed.  Unlike nonzero temperature --- where
only the $\sim \alpha_s^3(\log^2(\alpha_s),\log(\alpha_s))$
are sensible to compute, for dense QCD {\it all}
terms in the perturbative expansion of the thermodynamic potential
are well defined.

This does not imply that a given
value of $\alpha_s$, which is adequate to compute the thermodynamic
potential, works equally well for all other quantities.  In particular,
the gaps for color superconductivity are nonperturbative,
$\phi \sim \exp(-1/\sqrt{\alpha_s})$ \cite{son,superreview},
and much smaller values of
$\alpha_s$ appear to be required to reliably compute them
\cite{Rajagopal00}.  Perturbative results,
unjustly extrapolated down to 
$\mu \sim 400$~MeV, suggest that these gaps are  
$\sim 30$~MeV \cite{dirk:statisticalQCD}. 
As the relative change in the thermodynamic potential is only
$\sim (\phi/\mu)^2$, then, for the equation of state in QCD, 
color superconductivity is a negligible effect.

\begin{figure}[htb]
\begin{minipage}[t]{71mm}
\includegraphics[width=7cm]{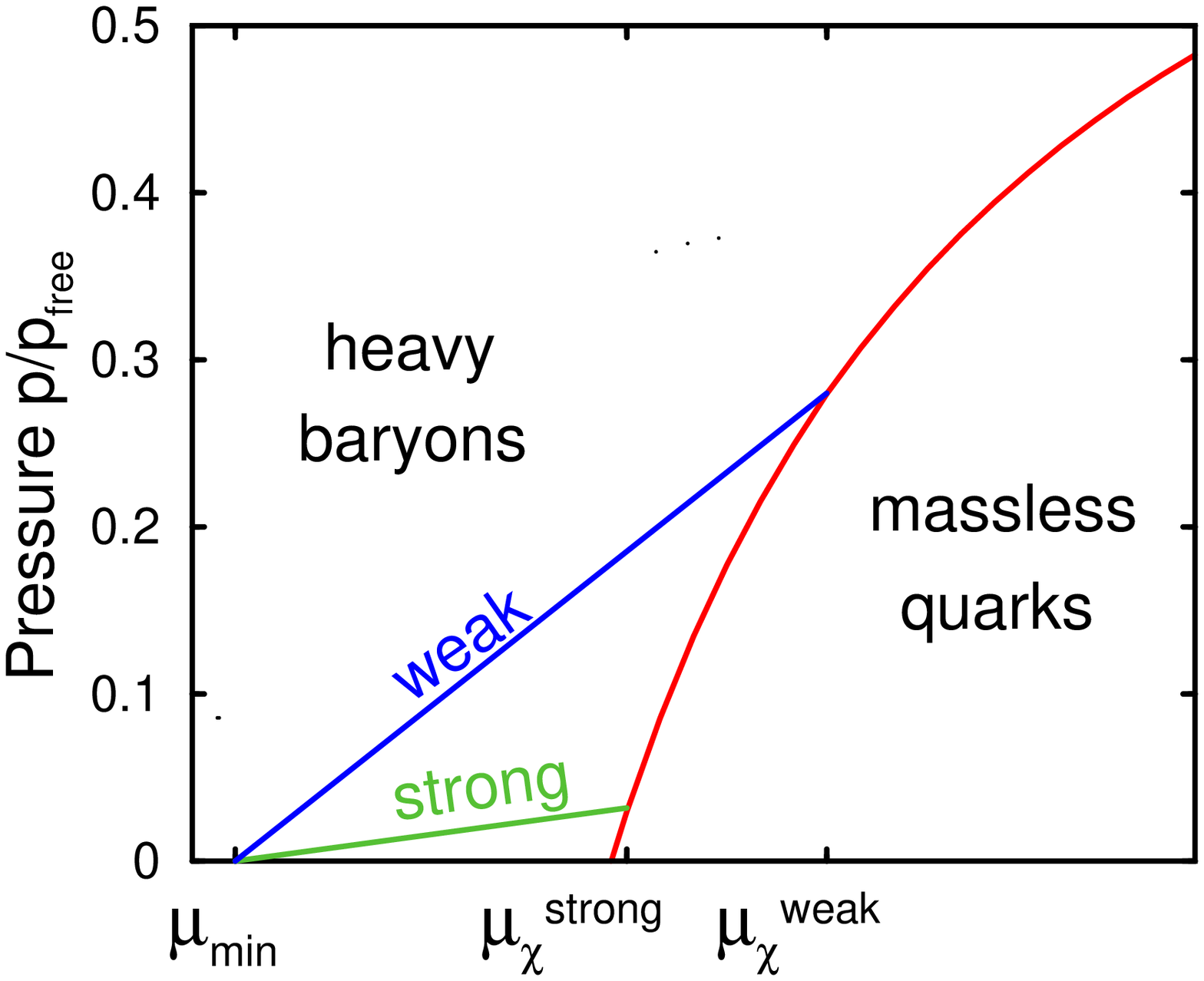}
\caption{Sketch of the total pressure, relative to the pressure of an
ideal gas, $p_{free}$, for a weak (upper line) and a strong (lower line) chiral 
transition as discussed in the text.}
\label{fig:scenarios}
\end{minipage}
\hspace{\fill}
\begin{minipage}[t]{71mm}
\includegraphics[width=7cm]{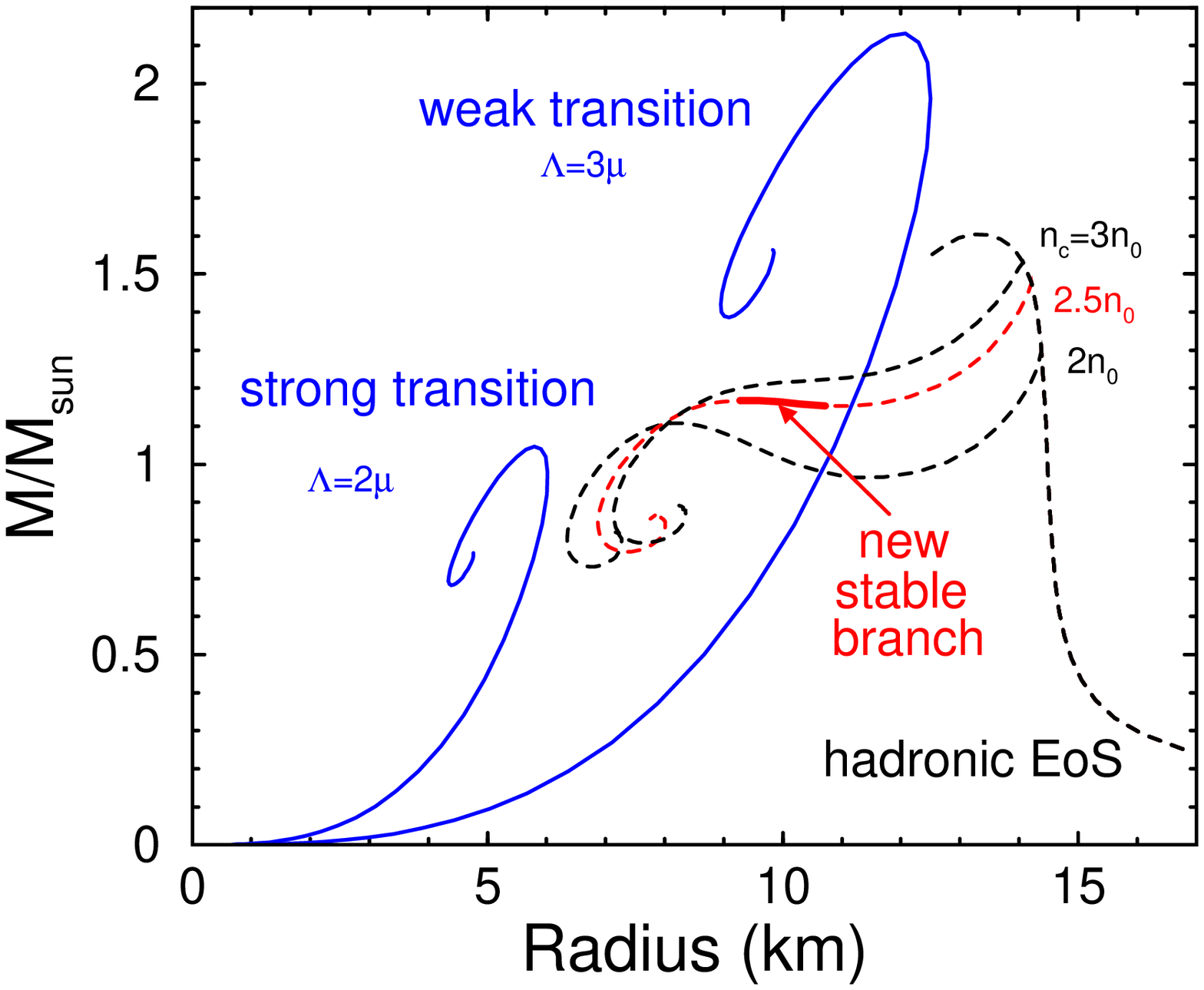}
\caption{Mass-radius relation 
for pure quark stars (solid lines) and hybrid stars (dashed lines) for
different critical densities for the chiral transition. 
}
\label{fig:mr_quarkhadron}
\end{minipage}
\end{figure}

The structure of a quark star is determined by the solution to the
Tolman-Oppenheimer-Volkov (TOV) equations \cite{G_b}. 
Given the equation of state $p=p(\epsilon)$, 
one can integrate the TOV equations from the origin until the pressure 
vanishes at the edge of the star, obtaining a characteristic
curve in the mass-radius diagram.
In matching the pressure of quarks onto that of hadrons, there
are two possible scenarios.
In the first, the pressure of hadrons rise quickly, 
smoothly matching onto that of quarks.
Since the pressure of the hadronic and quark phases are similar,
so are the densities (as the slopes of the pressure);
then
the chiral phase transition is either weakly first order, or just crossover.  
In the second case, the hadronic pressure is assumed to always
remain small.
This gives a strongly first order chiral transition, as one goes
abruptly from a phase of dilute hadrons to one of dense quarks \cite{strong}.
A sketch of both cases is depicted in Fig. \ref{fig:scenarios}, where 
the equation of state for massless quarks is taken to be 
the one with $\bar\Lambda=2\mu$.

Figure \ref{fig:mr_quarkhadron} shows 
the mass-radius diagram for these two cases.
A weakly first order chiral transition is illustrated by 
the choice of $\bar\Lambda=3\mu$.  The quark pressure vanishes
at $\mu_c = \mu_{min}$, so that the pressure of dense hadrons
grows immediately, and becomes equivalent to the pressure of
dense quarks.   The numerical solution of the TOV equations for such 
an equation of state generates hybrid stars with maximum masses and radii
compatible with those from conventional neutron star models,
$M \sim 2\, M_\odot$ and $R \sim 12$~km. 

For a strongly first order chiral transition, as with
$\bar\Lambda = 2 \mu$, there is a branch along which
stars are predominantly composed of dilute hadrons, with
small pressure.  These look like ordinary neutron stars.
In addition, there is a new branch in the mass-radius diagram.
These are predominantly composed of very dense quarks with a thin
hadronic mantle.  The pressure is like that of QCD with a larger
bag constant ($\mu_{min}$ is larger).  
Working out how the maximum mass and radius scale
with an (effective) bag constant, one finds maximum masses
$M \sim 1\, M_\odot$ and radii $R \sim 6$~km. 
The new branch was suggested previously by 
several groups \cite{thirda,thirdb}.

Since we have no plausible hadronic equation of state with small
pressure, we simply ignore it, and take the quark equation of
state down to the point where it vanishes.  In comparison,
in Figure \ref{fig:mr_quarkhadron} we also show the mass-radius
relation for a hadronic equation of state
(a mean-field equation of state with parameter set TM1, including hyperons 
\cite{sm}). 
Matching the hadronic equation of 
state onto the one for the quark phase is done by fixing a critical density. 
For critical densities about twice that of nuclear matter, we find 
a new branch appears, characterized by a 
second bump in the mass-radius curve.  

Experimentally, Pons {\it et al} \cite{Pons:2001px}
find that for an isolated, nearby neutron star, the
masses and radii are {\it exactly} what one expects for
a new branch of quark stars, with 
$M \approx .9 M_\odot$ and $R \approx 6$~km.

The difficulty is really in understanding how hadrons
can have a small pressure \cite{soft}.
Akmal, Pandharipande, and Ravenhall \cite{Akmal1998} have
found that to a very good approximation, in pure neutron matter
the energy per baryon (minus its rest mass)
is approximately linear in the baryon density, $n$:
\begin{equation}
\frac{E}{A} -m = \frac{\epsilon}{n} - m = 15 \; {\rm MeV} \left(\frac{n}{n_0}\right) \; .
\label{en}
\end{equation}
where $\epsilon$ is the energy density and
$n_0 \sim .16$ baryons/fm$^3$ is the saturation density for nuclear matter.
These authors used a wide variety of different models for the 
nucleon-nucleon interaction, and find that up to $n \approx n_0$,
all models are in close agreement.  Even at such
``low'' densities, this behavior itself is already
signs of a {\it highly} nonideal Fermi liquid, since free fermions
give $\epsilon/n - m \sim n^{2/3}$, not $\sim n$.  By two times nuclear
matter density, the discrepancies between different models is
already $\sim 50$\%, and grow sharply with density.

The energy on the right hand side of (\ref{en}), $\sim 15$~MeV, is 
very small on hadronic scales.  
It is much smaller than the pion decay constant,
which one might expect is the natural scale.  
What is remarkable about the results of Akmal {\it et al.} \cite{Akmal1998}
is that the energies are small not only
for nuclear matter, which has a nonzero binding energy,
but even for pure neutron matter, which is unbound.
This suggests to us that the smallness of the right hand
side of eq.~(\ref{en}) is a generic property of baryons interacting
with pions, {\it etc.}, and is not due to any special tuning.
Using this result, however, we can reliably compute the pressure
of pure neutron matter at nuclear matter density:
\begin{equation}
p_{hadron} = n^2 \frac{\partial}{\partial n} 
\left(\frac{\epsilon}{n} \right) \; = \;
15  \; {\rm MeV} \; \frac{n^2}{n_0} \; = \; .04 \;\; p_{free}  \; 
\left(\frac{n}{n_0}\right)^2
\end{equation}
Here $p_{free}$ is the pressure for three flavors of massless quarks;
for the purposes of discussion, in $p_{free}$ we have taken 
$\mu = 313$~MeV.  Thus at nuclear matter density, the pressure
is $4 $\% that for ideal quarks, which is comfortably small.
For the hadronic pressure to remain small relative to the quark
pressure, the transition must occur at remarkably low densities,
maybe twice that of nuclear matter.

We do not have a good reason for why the hadronic pressure should
remain small to the chiral transition.  But since we don't understand
why it is small in nuclear matter, we feel free to extrapolate;
maybe the hadronic pressure
is suppressed in the limit of a large number of colors.
Such ``implausible'' equations of state, with a small hadronic
pressure, can be constructed mathematically \cite{progress}.

If small and dense quark stars occur, then their production is probably
different
from those of ordinary neutron stars. Models of supernova collapses suggest 
that the mass of a proto-neutron star formed by the collapse of the iron core 
is in excess of 1$M_\odot$ as the initial mass of the iron core is well above 
1$M_\odot$. Neutron stars can not collapse to quark stars, as they have to 
expel most of their mass to achieve a stable quark star. 
White dwarfs do not collapse to more compact objects in type Ia supernova 
as the explosion destroys the whole star.
In the early universe, blobs of strange quark matter formed 
during the quark-hadron phase transition can not be larger than 
$10^{-9}M_\odot$ due to the horizon at $t=10^{-6}$s after the big-bang
\cite{madsen}. 

Nevertheless, there might be some exotic ways to 
produce quark stars.
If white dwarfs can develop an iron core by some (so far not established) 
reason, the collapse of this core can form a low-mass quark star.
Alternately, if inflation occurs at the quark-hadron phase transition,
then larger strange quark  
stars, with $M=0.01-10M_\odot$, can be formed \cite{inflation}.   

The existence of a new class of small and dense compact stars is an
exciting possibility.  While superconducting condensates
do not matter for the pressure, they do dominate their
electromagnetic response.  Whatever the magnitude of
the gaps for color superconductivity, versus those for hadrons,
there should be a multitude of
ways in which stars composed of color-flavor locked quarks
differ from those of superconducting (and superfluid) baryons.
After all, in this scenario the stars are never really hybrids,
but either mainly hadronic, or mainly quark.

 

We thank G. Baym, J. Beacom, H.-T. Janka, A. Olinto, A. Peshier, K. Rajagopal, 
D. Rischke, and especially L. McLerran for fruitful discussions.
We thank the U.S. Department of Energy for their support under Contract
No. DE-AC02-98CH10886; E.S.F., for the support of CNPq (Brazil);
and J.S.B., for the support of RIKEN and BNL.



\begin{thebibliography}{9}
\bibitem{soft}
When $\mu < \mu_\chi$, the pressure is small if it 
is dominated by the thermal Polyakov loop 
(R.~D.~Pisarski, Phys.\ Rev.\ D {\bf 62}, 111501 (2000)),
and if its expectation value is also small.
Why this should be true, 
especially if $N_f \approx N_c$, is unclear \cite{dum}.  Perhaps
$m_s \neq 0$ has a greater effect when $\mu < \mu_\chi$ (kaons
are much heavier than pions) than for $\mu > \mu_\chi$.
\end{thebibliography}
\end{document}